\documentclass[prb,twocolumn]{revtex4}

\usepackage{graphicx}
\usepackage{dcolumn}
\usepackage{bm}

\begin{document}


\title{Photon correlation studies of single GaN quantum dots}

\author{Charles Santori}
\email{cmsantori@yahoo.com}
\author{Stephan G\"{o}tzinger}
\author{Yoshihisa Yamamoto}
\affiliation{E.L. Ginzton Laboratory, Stanford University, Stanford, CA 94305}
\author{Satoshi Kako}
\author{Katsuyuki Hoshino}
\author{Yasuhiko Arakawa}
\affiliation{Institute of Industrial Science, University of Tokyo,
4-6-1 Komaba, Meguro-ku, Tokyo 153-8505, Japan}
\date{\today}

\begin{abstract}
We present measurements of the second-order coherence function on emission from
single GaN quantum dots.  In some cases a large degree of photon antibunching
is observed, demonstrating isolation of a single quantum system.  For a selected
quantum dot, we study the dependence of photon antibunching on excitation power
and temperature.  Using pulsed excitation, we demonstrate an ultraviolet triggered
single-photon source operating at a wavelength of 358~nm.
\end{abstract}

\maketitle

Quantum dots~\cite{arakawa} (QDs) in nitride semiconductors, potentially
useful for improving the performance of short-wavelength devices such as
light-emitting diodes and lasers~\cite{nakamura}, also offer exciting possibilities
for optical quantum information applications.  For example, optically addressed
excitonic qubits could exhibit large two-qubit interactions due to the large built-in
electric field found in nitride QDs~\cite{rinaldis}.  In addition, with a wide range of
possible emission wavelengths, nitride QDs are interesting as single-photon sources
for applications such as free-space quantum cryptography~\cite{rarity}, where a shorter
wavelength could in principle allow for smaller
transmitter and receiver telescopes.  Another motivation is the potential for
higher-temperature devices.  Although II-VI compounds have been investigated
for this purpose~\cite{sebald}, nitride semiconductors are more resistant to
degradation, and are suitable for electrically-contacted devices.

Recently, Stranski-Krastanov growth of high-quality GaN QDs on AlN has been demonstrated
by metalorganic chemical vapor deposition~\cite{miyamura}.  Photoluminescence from single
GaN QDs has been reported, and discrete spectral peaks have been identified that exhibit
excitation power dependencies consistent with exciton and biexciton
behavior~\cite{kako04}.  Here, we report measurements of the second-order coherence
function~\cite{walls}
\mbox{$g^{(2)}(\tau) = \langle:I(0)I(\tau):\rangle / \langle I \rangle^2$}
on single GaN QDs.  For some QDs, strong nonclassical behavior (photon antibunching) is
observed, demonstrating isolation of a single anharmonic quantum system.
Occasionally, we instead observe bunching on a nanosecond timescale.
For a selected QD, we study the dependence of photon antibunching on excitation power and
temperature, and demonstrate triggered single-photon generation.

For this study, hexagonal GaN/AlN QDs were grown
on a (0001)-oriented 6H-SiC substrate.  QDs with an average height of approximately 4~nm, an
average diameter of 20~nm and a density of approximately $10^{10}\,{\rm cm}^{-2}$ were grown
on top of a 200~nm AlN buffer layer and covered with a 50~nm AlN cap layer.  To isolate small
numbers of QDs, mesas with diameters from $0.2 - 2\,\mu{\rm m}$ were fabricated by electron-beam
lithography. 
\begin{figure}
\includegraphics{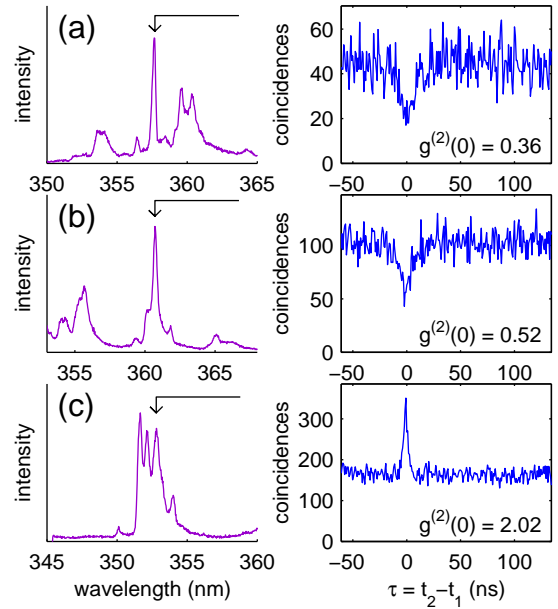}
\caption{\label{fig1} (a)-(c) Left: Photoluminescence spectra of mesas A, B and C,
respectively.  The arrows indicate peaks selected for photon correlation measurements on
right.  Right: Number of measured coincidences vs. relative delay $\tau$ between two detectors.
The time resolution (bin size) is 0.81~ns.  Total count rates on two detectors:
(a) 8200~$s^{-1}$, (b) 8000~$s^{-1}$, (c) 13300~$s^{-1}$; temperatures $<20{\rm K}$.}
\end{figure}

The sample was cooled to temperatures from 10-100K in a liquid-helium continuous-flow
cryostat providing optical access through a thin window.  The QD
wetting layer was excited at a wavelength of 266~nm using a beam produced by second-harmonic
generation with a frequency-doubled Nd:YVO$_4$ laser.  This beam was focused from a steep
angle to a $\sim 20\,\mu {\rm m}$ spot on the sample.  The resulting photoluminescence was
collected using an NA=0.6 microscope objective with cover-slip correction,
and imaged onto a pinhole that selected
a $2\,\mu {\rm m}$ region on the sample.  The light was then sent to a cooled-CCD
spectrometer or to a Hanbury Brown-Twiss (HBT) photon correlation setup.
For the HBT setup, the light was first spectrally filtered using a prism-based monochromator
configuration with a transmission bandwidth adjustable to $< 1\,{\rm nm}$.
The light was then split into two paths by a beamsplitter, each path leading to a
miniature photomultiplier-tube detector (200~ps time resolution, $50\,{\rm s}^{-1}$ dark
counts, $20 \%$ efficiency).  A time-to-amplitude converter followed by a
multi-channel-analyzer computer card produced a histogram of the relative
delay $\tau = t_2-t_1$ between photon detections on counters 1 and 2.  Integration
times were typically at least one hour.  A computer-controlled feedback system
prevented the sample position from drifting during this time.

Spectroscopy of single mesas showed a wide variety of patterns of spectral peaks.  Individual
peaks typically showed a large degree of linear polarization along various directions.  The
spectral linewidths were highly variable.  Most of the peaks showed at least a small amount
of spectral diffusion on timescales of seconds or longer, and some peaks blinked.  The results
presented here are from QDs in $0.4\,\mu {\rm m}$ mesas selected for bright emission, narrow
peaks and relatively stable spectra.
\begin{figure}
\includegraphics{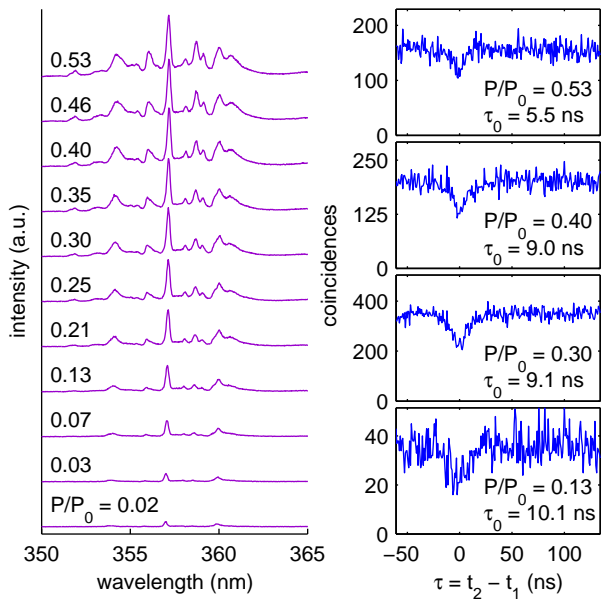}
\caption{\label{fig2} Left: spectra measured at various excitation
powers.  Normalized powers $P/P_0$ are indicated.  Right: photon
correlation measurements at selected powers.}
\end{figure}

Figure~1 (left) shows photoluminescence spectra from three mesas labeled
A, B, and C obtained using excitation powers well below saturation.  The corresponding
photon correlation data (right) were fitted with the function
$G^{(2)}(\tau) = A \left[1 - (1-g^{(2)}(0)) \exp(-|\tau|/\tau_0)\right]$
to obtain the fitting parameters $A$, $g^{(2)}(0)$ and $\tau_0$.  When $g^{(2)}(0)=0$,
this describes a simple two-level system~\cite{michler} with a spontaneous
emission rate $\Gamma$, pump rate $r$, and $1/\tau_0 = \Gamma + r$.
Mesa~A produced a well-isolated sharp line (FWHM = 0.23~nm) at 357.6~nm.  The correlation
measurement shows a large degree of photon antibunching with fitting parameter
$g^{(2)}(0)=0.36$.  For mesa~B, shoulders appear connected to the main emission
line at 360.7~nm.  Antibunching was observed when the spectral
filter was closed narrowly around the main peak, but the antibunching vanished when the
filter was adjusted to include the shorter-wavelength emission, which is likely biexcitonic in origin.
Mesa~C produced a group of three bright spectral peaks.  When the longest-wavelength peak
was selected, strong photon bunching was observed with $g^{(2)}(0) \approx 2$.  When the excitation
power was increased, the correlation peak became shorter and narrower.  Positive
correlations with nanosecond timescales indicate a two-photon cascade
process~\cite{moreau} that could occur, for example, if
both detectors simultaneously measure exciton and biexciton luminescence.
\begin{figure}
\includegraphics{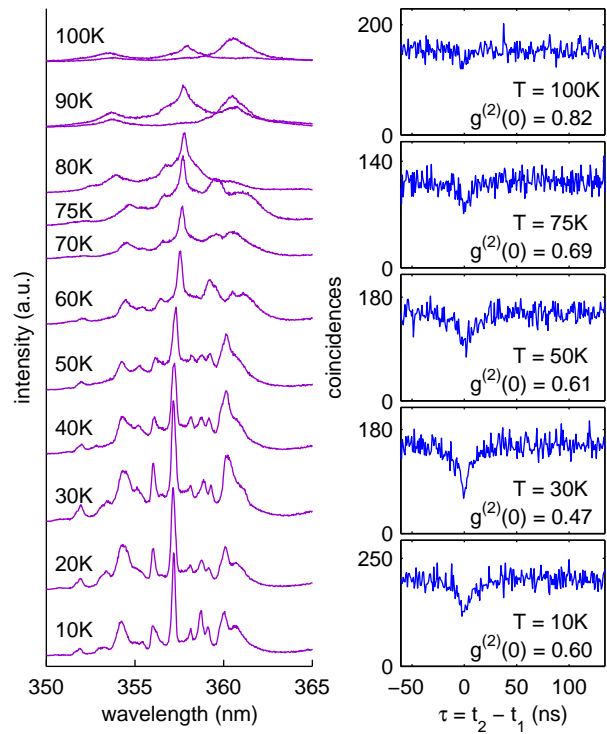}
\caption{\label{fig3} Left: spectra measured at various temperatures
as indicated.  At 90K and 100K, blinking behavior was observed, and two
representative spectra are plotted.  Right: photon correlation measurements at
selected temperatures.}
\end{figure}

The excitation power dependence for mesa~A is shown in Fig.~2.  The normalized excitation
powers $P/P_0$ are indicated, with the parameter $P_0 \approx 10^3 \, {\rm W\,cm}^{-2}$
obtained by fitting the intensity of the main emission peak with a Poisson distribution
model for the single-exciton recombination intensity, $I/I_0=(P/P_0)\exp(-P/P_0)$.
A small peak at slightly
shorter wavelength increases nonlinearly with power and may be a biexciton peak.
On the right, photon correlation measurements obtained at selected excitation powers
are shown, and the fitted antibunching timescale $\tau_0$ is indicated in each case.
The timescale changes little at weak excitation powers, but changes rapidly
at higher powers.  Qualitatively, this agrees with the two-level model above,
where $\tau_0 = 1/(\Gamma + r)$ converges to $1/\Gamma$ in the weak-excitation limit.
At weak excitation power, $\tau_0 \approx 9\,{\rm ns}$, in good agreement
with the spontaneous emission lifetime $1/\Gamma = 8\,{\rm ns}$ which we have
measured independently using a pulsed excitation source.

The temperature dependence for mesa~A from $10-100{\rm K}$ is shown in Fig.~3.
As the temperature is increased, at first only a background appears to grow in the spectra,
but by $50-60{\rm K}$ line broadening appears, possibly from acoustic phonon
sidebands~\cite{besombes}.  This broadening increases rapidly as the temperature increases
further.  At $90-100{\rm K}$, the spectra began to blink with a timescale on the order of
$1-10\,{\rm s}$.  Two representative spectra from different blinking configurations are
shown in these cases.  In the photon correlation measurements, the spectral filter width was
kept constant, but the center wavelength was adjusted as necessary to follow the main
emission line.  The antibunching decreases only slightly from $10-50{\rm K}$, but fades rapidly from
$50-100{\rm K}$, reflecting the broadening in the spectra.

Finally, we used pulsed excitation of mesa~A to demonstrate an ultraviolet triggered
single-photon source.  For such a source, the probability
of measuring more than one photon in a given pulse is reduced compared to a Poisson distribution with
the same mean photon number.  Fig.~4 shows a photon correlation measurement at 10K of mesa~A
excited with 266~nm excitation pulses, obtained through third-harmonic generation
with a Ti-sapphire laser.  The pulses had a duration of $\sim 2\,{\rm ps}$, and the repetition
period was reduced to~78.6~ns using a pulse picker.  The measurement shows a series
of peaks, with the peak at $\tau=0$ corresponding to events where two photons were detected
following the same excitation pulse.  When detector dark counts are taken into account,
the reduced area of the central peak corresponds to a reduction of the two-photon
probability to $0.24 \pm 0.03$ times that for an equivalent Poisson-distributed source.
\begin{figure}
\includegraphics{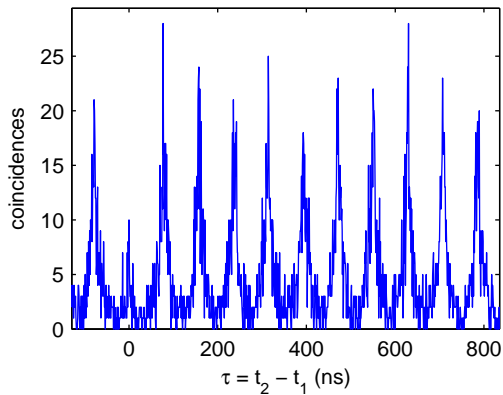}
\caption{\label{fig4} Photon correlation measurement of mesa~A under pulsed excitation.
The total count rate was 1400~$s^{-1}$, and the time resolution (bin size) is $0.97\,{\rm ns}$. }
\end{figure}

These results constitute the first demonstration of photon antibunching and triggered
single-photon generation at ultraviolet wavelengths using GaN quantum dots.  Devices capable of
operating at higher temperatures should be possible if the exciton-biexciton spectral separation
can be increased through modification of the quantum-dot size and composition~\cite{oreilly}.
Another goal will be to decrease the long spontaneous emission lifetimes observed
in GaN quantum dots, which have been attributed to built-in electric fields that cause
spatial separation of the electron and hole wavefunctions~\cite{daudin,kako03}.
Elimination or reduction of the built-in field is therefore a priority for
improving GaN-based single-photon sources.  This may also help to decrease the spectral
bandwidth, since the built-in field is thought to enhance the interaction
with fluctuating electric fields~\cite{kako04}.

This work is supported by IT Program MEXT, JST SORST program on quantum
entanglement, NTT Basic Research Laboratories, and ARO MURI \#DAAD19-03-1-0199-P00001.
S. G\"{o}tzinger is supported by the Alexander von Humboldt foundation.

\end{document}